\documentclass[12pt]{article}
\usepackage{amsfonts,bm}
\usepackage{graphicx}
\baselineskip
\offinterlineskip
\begin{document}

\begin{center}
{\bf Exceptional Points, Nonnormal Matrices, Hierarchy of Spin 
Matrices and an Eigenvalue Problem}
\end{center}

\begin{center}
{\bf  Willi-Hans Steeb$^\dag$ and Yorick Hardy$^\ast$} \\[2ex]

$\dag$
International School for Scientific Computing, \\
University of Johannesburg, Auckland Park 2006, South Africa, \\
e-mail: {\tt steebwilli@gmail.com}\\[2ex]

$\ast$
Department of Mathematical Sciences, \\
University of South Africa, Pretoria, South Africa, \\
e-mail: {\tt hardyy@unisa.ac.za}\\[2ex]
\end{center}

\strut\hfill

{\bf Abstract} Exceptional points of a class of non-hermitian Hamilton 
operators $\hat H$ of the form $\hat H=\hat H_0+i\hat H_1$ are studied,
where $\hat H_0$ and $\hat H_1$ are hermitian operators.
Finite dimensional Hilbert spaces are considered.
The linear operators $\hat H_0$ and $\hat H_1$ are given by spin matrices
for spin $s=1/2,1,3/2,\dots$. Since the linear operators studied
are nonnormal, properties of such operators are described.

\strut\hfill

\section{Introduction}

Non-hermitian Hamilton operators in finite dimensional
Hilbert spaces of the form $\hat H=\hat H_0+i\hat H_1$
have been studied by several authors
(Alcaraz et al \cite{1}, Henkel and Herrmann \cite{2} and
Brody \cite{3}). Here $\hat H_0$ and $\hat H_1$ are hermitian operators.
If the commutator of $\hat H_0$ and $\hat H_1$ does not vanish,
then the linear operator $\hat H$ is nonnormal.
Here we study exceptional points for such Hamilton operators,
where $\hat H_0$ and $\hat H_1$ are given by spin matrices
for spin $s=1/2,1,3/2,\dots$.
Kato \cite{4} (see also Rellich \cite{5}) introduced exceptional 
points for singularities appearing in the perturbation theory of 
linear operators. Afterwards exceptional points and energy level crossing have been studied
for hermitian Hamilton operators \cite{6,7,8,9,10,11,12,13}
and non-hermitian Hamilton operators \cite{14,15,16,17,18,19}
by many authors. Here we consider the finite dimensional Hilbert 
space ${\mathbb C}^n$ and the linear operators are $n \times n$
matrices over $\mathbb C$. 
\newline

For hermitian matrices the standard example in literature is
$$
H(\epsilon) = \pmatrix { 0 & 0 \cr 0 & 1 } +
\epsilon \pmatrix { 0 & 1 \cr 1 & 0 }
$$
where $\epsilon$ is real. The characteristic polynomial 
$\det(H(\epsilon)-EI_2)=0$ is given by $E^2-E-\epsilon^2=0$. 
When $\epsilon$ is complex, the eigenvalues may be viewed 
as the $2$ values of a single function $E(\epsilon)$ of
$\epsilon$, analytic on a Riemann surface with 2 sheets 
joined at branch point singularities in the complex plane.
The exceptional points in the complex $\epsilon$ plane
are defined by the solution $\det(H(\epsilon)-EI_2)=0$
together with $d(\det(H(\epsilon)-EI_2))/dE=0$. 
One finds that the exceptional points are $\epsilon_1=i/2$ and
$\epsilon_2=-i/2$. 
\newline

For non-hermitian systems the standard example is the matrix
(Kato \cite{4}, Rotter \cite{14}, Heiss \cite{15}) 
$$
\sigma_3 + z\sigma_1 = \pmatrix { 1 & z \cr z & -1 } 
$$
where $z \in {\mathbb C}$ and $\sigma_1$, $\sigma_2$, $\sigma_3$ are
the Pauli spin matrices
$$
\sigma_1 = \pmatrix { 0 & 1 \cr 1 & 0 }, \quad
\sigma_2 = \pmatrix { 0 & -i \cr i & 0 }, \quad
\sigma_3 = \pmatrix { 1 & 0 \cr 0 & -1 }.
$$
Let $z=i$. Then the matrix $\sigma_3+i\sigma_1$ admits the eigenvalue 0 
(twice) and the only normalized eigenvector
$$
{\bf v} = \frac1{\sqrt2} \pmatrix { -i \cr 1 }.
$$
The matrix $\sigma_3+i\sigma_1$ is nonnormal.
Let $z=-i$. Then the nonnormal matrix $\sigma_3-i\sigma_1$ admits the 
eigenvalue 0 (twice) and the only normalized eigenvector
$$
{\bf w} = \frac1{\sqrt2} \pmatrix { i \cr 1 }
$$
where ${\bf v}^*{\bf w}=0$. We extend this result to arbitrary spin. 
Since the matrices considered are nonnormal we summarize
the properties of nonnormal matrices in section 2.
In section 3 we consider the cases with spin $1/2$, $1$, $3/2$ 
and $2$. In section 4 the general case is studied.
The spectra of the non-hermitian operators are provided.

\section{Nonnormal Matrices}

An $n \times n$ matrix $A$ over $\mathbb C$ is called normal
if $AA^*=A^*A$. Then for a nonnormal matrix we have 
$A^*A \ne AA^*$. An example of a nonnormal matrix is 
the matrix given above $\sigma_3+i\sigma_1$
which only admits the eigenvalue 0 (twice) and only one
eigenvector. Note that not all nonnormal matrices are 
non-diagonalizable, but all non-diagonalizable matrices are
nonnormal \cite{20}. 
\newline

If $A$ is any $n \times n$ matrix $A$ over $\mathbb C$, then a classical 
result due to Schur (Roman \cite{21}, Steeb \cite{22})
states that there exist a unitary matrix $U$ and a triangular
matrix $T=(t_{jk})$ with $t_{jk}=0$ for $k < j$ such that $A=UTU^*$.
For the matrix $\sigma_3+i\sigma_1$ we find (Schur decomposition)
$$
\sigma_3 + i\sigma_1 = \frac1{\sqrt2} \pmatrix { 1 & i \cr i & 1 }
\pmatrix { 0 & 2i \cr 0 & 0 } 
\frac1{\sqrt2} \pmatrix { 1 & -i \cr -i & 1 }.
$$
Let $A$, $B$ be hermitian nonzero matrices, i.e. $A^*=A$ and $B^*=B$.
Consider the matrix $A+iB$. What are the conditions on $A$ and $B$ 
such that $A+iB$ is normal? From
$$
(A+iB)^*(A+iB) = (A+iB)(A+iB)^*
$$
we find that the commutator of $A$ and $B$ must vanish, i.e.
$[A,B]=0$. For the matrix $\sigma_3+i\sigma_1$ this 
condition is not satisfied since $[\sigma_3,\sigma_1]=2i\sigma_2$.
Now the transition from a hermitian matrix to a non-normal
matrix can be studied with the matrix
$$
\sigma_3 + e^{i\phi}\sigma_1
$$
where $\phi \in [0,\pi/2]$. For $\phi=0$ we have the hermitian
matrix $\sigma_3+\sigma_1$. For $0 < \phi \le \pi/2$ we have 
a nonnormal matrix. The eigenvalues are given by
$$
\lambda_\pm = \pm \sqrt{1+e^{2i\phi}}
$$
with the eigenvectors
$$
{\bf v}_{\pm} = \pmatrix { e^{i\phi} \cr -1 + \lambda_{\pm} }.
$$ 
Note that the commutator of
$\sigma_3+\sigma_1$ and $\sigma_3+e^{i\phi}\sigma_1$ is given by
$$
[\sigma_3+\sigma_1,\sigma_3+e^{i\phi}\sigma_1] = 
2i\sigma_2(e^{i\phi}-1)\,.
$$
Obviously for $\phi=0$ the commutator vanishes and for 
$\phi=\pi/2$ we have $2i\sigma_2(i-1)$. The matrix 
$2i\sigma_2(i-1)$ is normal, but non-hermitian.
\newline

Measures of nonnormality of square matrices have been discussed
by Henrici \cite{23} and Elsner and Paardekooper \cite{24}.
A natural measure of nonnormality of a square matrix $A$ is
$$
m(A) = \|A^*A-AA^*\|_2.
$$
With the example $A=\sigma_3+e^{i\phi}\sigma_1$ and $\phi \in [0,\pi/2]$ 
we obtain $m(A)=4|\sin(\phi)|$.
\newline

Let $\otimes$ be the Kronecker product and $\oplus$ the direct sum.
Let $A$, $B$ be nonnormal matrices. Then $A \otimes B$ and
$A \oplus B$ are nonnormal. Let $X$, $Y$ be non-zero 
$n \times n$ matrices. We have
$$
(X^*X) \otimes (Y^*Y) = (XX^*) \otimes (YY^*)
$$
if and only if $X^*X=XX^*$ and $Y^*Y=YY^*$. Note that
$$
\exp(\sigma_3+i\sigma_1) = I_2 + \sigma_3 + i\sigma_1.
$$
This matrix is nonnormal, but of course invertible. 
It is an element of the Lie group $SL(2,{\mathbb R})$.
However, we cannot conclude in general
that $\exp(A)$ of a nonnormal matrix $A$ is nonnormal.
Consider, for example, the matrix
$$
A = \pmatrix { i\pi & b \cr 0 & -i\pi }
$$
with $b \ne 0$. Then $\exp(A)$ is a normal matrix. However, 
if a matrix $M$ is nonnormal and nilpotent, then $\exp(M)$ is 
nonnormal. If $N$ is a normal matrix, then $\exp(N)$ is a normal matrix.

\section{Spin-$\frac12$, $1$, $3/2$, $2$ Cases}

For the spin-$\frac12$ case we consider the spin matrices for 
describing a spin-$\frac12$ system
$$
s_1 = \frac12\sigma_1, \quad 
s_2 = \frac12\sigma_2, \quad
s_3 = \frac12\sigma_3 
$$
with $s_1^2+s_2^2+s_3^2=\frac34 I_2$.
Consider the matrix $s_3+is_1$. This is the case given above
except for the factor $1/2$.
Obviously the matrix $s_3+is_1$ is nonnormal and the rank is 1.
Since $(s_3+is_1)^2=0_2$ the matrix is nilpotent and thus
the eigenvalues are 0. The trace of this nonnormal matrix is 0. 
The eigenvalues of the matrix are 0 (twice) and only normalized 
eigenvectors of the matrix is 
$$
\frac1{\sqrt2} \pmatrix { -i \cr 1 }.  
$$
Consider next the spin matrices for describing a spin-$1$ system
$$
s_1 = \frac1{\sqrt2} 
\pmatrix { 0 & 1 & 0 \cr 1 & 0 & 1 \cr 0 & 1 & 0 }, \quad
s_2 = \frac1{\sqrt2} 
\pmatrix { 0 & -i & 0 \cr i & 0 & -i \cr 0 & i & 0 }, \quad
s_3 = \pmatrix { 1 & 0 & 0 \cr 0 & 0 & 0 \cr 0 & 0 & -1 } 
$$
with $s_1^2+s_2^2+s_3^2=2I_3$. For spin-1 the matrix
$$
s_3 + is_1 = \pmatrix { 1 & i/\sqrt2 & 0 \cr i/\sqrt2 & 0 & i/\sqrt2 \cr
0 & i/\sqrt2 & -1 }
$$
is nonnormal. The trace of this nonnormal matrix is 0 and the matrix 
is nilpotent, i.e. we have $(s_3+is_1)^3=0_3$. Thus all three eigenvalues 
are 0 and the only normalized eigenvector is
$$
\frac12 \pmatrix { -1 \cr -i\sqrt2 \cr 1 }.
$$
For spin-$3/2$ we have the hermitian matrices
$$
s_1 = \frac12 \pmatrix { 0 & \sqrt3 & 0 & 0 \cr \sqrt3 & 0 & 2 & 0 \cr
                 0 & 2 & 0 & \sqrt3 \cr 0 & 0 & \sqrt3 & 0 }, \quad
s_2 = \frac12 \pmatrix { 0 & -i\sqrt3 & 0 & 0 \cr 
                 i\sqrt3 & 0 & -2i & 0 \cr
                 0 & 2i & 0 & -i\sqrt3 \cr 
                 0 & 0 & i\sqrt3 & 0 },
$$
$$
s_3 = \pmatrix { 3/2 & 0 & 0 & 0 \cr 0 & 1/2 & 0 & 0 \cr 
                 0 & 0 & -1/2 & 0 \cr 0 & 0 & 0 & -3/2 }
$$
with $s_1^2+s_2^2+s_3^2=\frac{15}{4}I_4$. Thus the matrix $s_3+is_1$
is given by 
$$
s_3 + is_1 = \pmatrix { 3/2 & i\sqrt3/2 & 0 & 0 \cr
           i\sqrt3/2 & 1/2 & i & 0 \cr
           0 & i & -1/2 & i\sqrt3/2 \cr
           0 & 0 & i\sqrt3/2 & -3/2 }.
$$
The matrix is nonnormal and nilpotent, i.e. $(s_3+is_1)^4=0_4$.
Thus the trace is equal to 0 and the eigenvalues are 0 (four times).
The rank of the matrix is 3. The only normalized eigenvector is
$$
\frac1{\sqrt8} \pmatrix { i \cr -\sqrt3 \cr -i\sqrt3 \cr 1 }.
$$
This eigenvector is entangled, i.e. it cannot be written as
a Kronecker product of two vectors in ${\mathbb C}^2$.
The tangle as a measure of entanglement is nonzero.
\newline

For spin-2 we have the hermitian $5 \times 5$ matrices
$$
s_1 = \pmatrix { 0 & 1 & 0 & 0 & 0 \cr 
                 1 & 0 & \sqrt6/2 & 0 & 0 \cr
                 0 & \sqrt6/2 & 0 & \sqrt6/2 & 0 \cr 
                 0 & 0 & \sqrt6/2 & 0 & 1 \cr
                 0 & 0 & 0 & 1 & 0 }, \quad
s_2 = \pmatrix { 0 & -i & 0 & 0 & 0 \cr 
                 i & 0 & -i\sqrt6/2 & 0 & 1 \cr
                 0 & i\sqrt6/2 & 0 & -i\sqrt6/2 & 0 \cr 
                 0 & 0 & i\sqrt6/2 & 0 & -i \cr
                 0 & 0 & 0 & i & 0 },
$$
$$
s_3 = \pmatrix { 2 & 0 & 0 & 0 & 0 \cr 0 & 1 & 0 & 0 & 0 \cr
                 0 & 0 & 0 & 0 & 0 \cr 0 & 0 & 0 & -1 & 0 \cr
                 0 & 0 & 0 & 0 & -2 }
$$
with $s_1^2+s_2^2+s_3^2=6I_5$. Thus the matrix $s_3+is_1$ is given by
$$
s_3 + is_1 = \pmatrix { 2 & i & 0 & 0 & 0 \cr
           i & 1 & i\sqrt6/2 & 0 & 0 \cr
           0 & i\sqrt6/2 & 0 & i\sqrt6/2 & 0 \cr
           0 & 0 & i\sqrt6/2 & -1 & i \cr
           0 & 0 & 0 & i & -2 }. 
$$
The matrix is nonnormal and nilpotent, i.e. $(s_3+is_1)^5=0_5$.
Thus the trace is equal to 0 and the eigenvalues are 0 (five times).
The rank of the matrix is 4. The only normalized eigenvector is
$$
\frac14 \pmatrix { 1 \cr 2i \cr -\sqrt6 \cr -2i \cr 1 }.
$$

\section{General Case}

For the general case we look at integer spin, i.e. $1,2,3,\dots$ 
and half-integer spin, i.e. $1/2,3/2,5/2,\dots$ \cite{19}.
Let $s$ (spin quantum number)
$$
s \in \left\{ \, \frac12, \, 1, \, \frac32, \, 2, \, \frac52, \dots
\,\, \right\}.
$$
Given a fixed $s$. The indices $j,k$ run over $s,s-1,s-2,\dots,-s+1,-s$.
Consider the $(2s+1)$ unit vectors (standard basis)
$$
{\bf e}_{s,s} = \pmatrix { 1 \cr 0 \cr 0 \cr \vdots \cr 0 }, \quad
{\bf e}_{s,s-1} = \pmatrix { 0 \cr 1 \cr 0 \cr \vdots \cr 0 }, 
\quad \ldots \,\,\, ,
{\bf e}_{s,-s} = \pmatrix { 0 \cr 0 \cr \vdots \cr 0 \cr 1 }.
$$
Obviously the vectors have $(2s+1)$ components. The
$(2s+1) \times (2s+1)$ matrices $s_+$ and $s_-$ are defined as
\begin{eqnarray*}
s_+ {\bf e}_{s,m} &:=&
\sqrt{(s-m)(s+m+1)} {\bf e}_{s,m+1}, \quad
m = s-1,s-2,\dots,-s \\
s_- {\bf e}_{s,m} &:=&
\sqrt{(s+m)(s-m+1)} {\bf e}_{s,m-1}, \quad
m = s,s-1,\dots,-s+1. 
\end{eqnarray*}
The $(2s+1) \times (2s+1)$ matrix $s_3$ is defined as
(eigenvalue equation)
$$
s_3{\bf e}_{s,m} := m{\bf e}_{s,m}, \quad m=s,s-1,\dots,-s.
$$
Thus $s_3$ is a diagonal matrix with the entries 
$s$, $s-1$, \dots, $-s$ in the diagonal.
Let ${\bf s}:=(s_1,s_2,s_3)$, where $s_+=s_1+is_2$ and 
$s_-=s_1-is_2$. Thus
$$
s_1 = \frac12 (s_+ + s_-), \qquad s_2 = -\frac{i}{2}(s_+ - s_-).
$$
We have
$$
(s_+)_{jk} = (s_-)_{kj} = \sqrt{(s-k)(s+k+1)}\delta_{j,k+1}
= \sqrt{(s+j)(s-j+1)}\delta_{j,k+1} 
$$
and
$$
(s_-)_{jk} = (s_+)_{kj} =
\sqrt{(s+k)(s-k+1)}\delta_{j,k-1}
= \sqrt{(s-j)(s+j+1)}\delta_{j,k-1}
$$
where $j,k=s,s-1,\dots,-s$. Therefore
$$
s_+ = 
\pmatrix { 0 & \sqrt{2s} & 0 & 0 & \dots & 0 \cr
           0 & 0 & \sqrt{2(2s-1)} & 0 & \dots & 0 \cr
           0 & 0 & 0 & \sqrt{3(2s-2)} & \dots & 0 \cr
        \vdots & \vdots & \vdots & \vdots & \ddots & \vdots \cr
           0 & 0 & 0 & 0 & \dots & \sqrt{2s} \cr 
           0 & 0 & 0 & 0 & \dots & 0 }.
$$
Thus $s_-=(s_+)^*$. We have $s_1^2+s_2^2+s_3^2=s(s+1)I_{2s+1}$
and $[s_1,s_2]=is_3$, $[s_2,s_3]=is_1$, $[s_3,s_1]=is_2$.  
Now the $(2s+1) \times (2s+1)$ matrix $s_3+is_1$ is nonnormal
and nilpotent, i.e.
$$
(s_3+is_1)^{2s+1} = 0_{2s+1}
$$
where $0_{2s+1}$ is the $(2s+1) \times (2s+1)$ zero matrix
and $s$ is the spin. Thus all the eigenvalues of 
$s_3+is_1$ are 0. For the eigenvectors $\bf v$ which is an element
of ${\mathbb C}^{2s+1}$ we set
$$
{\bf v} = \pmatrix { v_s & v_{s-1} & \cdots & v_{-s+1} & v_{-s} }^T.
$$
Now the eigenvalue equation $(s_3+is_1){\bf v}={\bf 0}$
can be easily solved. First we can set without loss
of generality the last entry of the eigenvector to $v_{-s}=1$.
Then using the last row of the matrix $s_3+is_1$ we obtain
the equation 
$$
i\frac12\sqrt{2s}v_{-s+1} - sv_{-s} = i\frac12\sqrt{2s}v_{-s+1}-s=0
$$
with the solution $v_{-s+1}=-i2s/\sqrt{2s}$.   
Then the second last row of the matrix $s_3+is_1$ provides
the equation for $v_{-s+2}$ and successively we can find
the other entries $v_{-s+3}, \dots, v_s$ of the eigenvector. 
All the entries are nonzero. This successive construction
also shows that there is only one linearly independent eigenvector.

\section{Conclusion}

We have studied an eigenvalue problem for a hierarchy of nonnormal 
matrices constructed from the spin matrices for spin $1/2$, $1$, 
$3/2$, $2$ etc. All these matrices have only one eigenvalue
and thus only one eigenvector. The matrices $(2s+1) \times (2s+1)$
matrices $s_3+is_2$ have the same properties as $s_3+is_1$, 
i.e. they are nonnormal and nilpotent for all spin $s$.
\newline

Starting from nonnormal matrices we can construct other
nonnormal matrices. Let $c_j^\dagger$, $c_j$ $(j=1,2)$ be Fermi 
creation and annihilation operators, respectively.
Then we can form the operator
$$
\pmatrix { c_1^\dagger & c_2^\dagger }
\pmatrix { 1 & i \cr i & -1 } 
\pmatrix { c_1 \cr c_2 } =
c_1^\dagger c_1 - c_2^\dagger c_2 
+ i(c_1^\dagger c_2 + c_2^\dagger c_1). 
$$
Using the four dimensional basis 
$|0\rangle$, $c_1^\dagger|0\rangle$, $c_2^\dagger|0\rangle$, 
$c_1^\dagger c_2^\dagger|0\rangle$ with $\langle 0|0\rangle=1$
we find the matrix representation
$$
\pmatrix { 0 & 0 & 0 & 0 \cr 0 & 1 & i & 0 \cr
           0 & i & -1 & 0 \cr 0 & 0 & 0 & 0 }
$$
for the operator. This matrix has the eigenvalue 0 (fourfold) and 
the three eigenvectors
$$
\pmatrix { 1 \cr 0 \cr 0 \cr 0 }, \quad
\frac1{\sqrt2} \pmatrix { 0 \cr 1 \cr i \cr 0 }, \quad
\pmatrix { 0 \cr 0 \cr 0 \cr 1 }.
$$
This matrix is nonnormal.
\newline

We can also consider the Kronecker product of such matrices.
Consider the case of spin-$\frac12$. The $4 \times 4$ matrix
$$
(\sigma_3+i\sigma_1) \otimes (\sigma_3+i\sigma_1)
$$
is nonnormal. However, note that the $4 \times 4$ matrix 
$$
\sigma_3 \otimes \sigma_3 + i \sigma_1 \otimes \sigma_1 =
\pmatrix { 1 & 0 & 0 & i \cr 0 & -1 & i & 0 \cr
           0 & i & -1 & 0 \cr i & 0 & 0 & 1 }
$$
is normal, but non-hermitian. The eigenvalues are
$$
(-1)^{1/4}\sqrt2, \quad -(-1)^{1/4}\sqrt2, \quad (-1)^{1/4}i\sqrt2, \quad 
(-1)^{1/4}i\sqrt2.
$$
\newline

{\bf Acknowledgment}
\newline

The authors are supported by the National Research Foundation (NRF),
South Africa. This work is based upon research supported by the National
Research Foundation. Any opinion, findings and conclusions or recommendations
expressed in this material are those of the author(s) and therefore the
NRF do not accept any liability in regard thereto.

\end{document}